\documentclass[11pt]{article}
\usepackage{graphicx}
\usepackage{amsmath}
\usepackage{amssymb}
\usepackage{amsxtra}

\title{Neutrinos production by photons scattered on the
	dark matter}
\author{V. Beylin\\Institute of Physics, SFedU, \\Stachki av. 194, Rostov-on-Don, Russia, 344090, vitbeylin@gmail.com}

\begin{document}

\maketitle

\begin{abstract}

In the framework of hypercolor scenario of multicomponent Dark Matter, inelastic interaction of high energy photons with the Dark Matter candidates is considered. 
This reaction results in production of energetic leptons and neutrinos, and the Dark Matter particles also can be boosted.
Total and differential cross sections have been calculated. The search of correlations between detected signals of high-energy photons and neutrinos can give an information on the Dark Matter scenario and dynamics. 
\end{abstract}


\section{Introduction}\label{s:intro}

Dark matter, which occupies such an important place in the observable Universe, both in its role in the formation of gravitating structures and in its contribution to the overall density of matter, still eludes the "hunters" - ground-based accelerators,  measuring complexes and underground laboratories do not find obvious signals of birth, decay or interactions of these objects of unknown nature. 
Extending the region of the "hunting", physicists are studying in more detail the indirect possible manifestations of Dark Matter in various astrophysical phenomena\cite{Feng,Roszkowski,Cirelli,Khlopov,Gaskins,Arcadi,Belotsky,Indirect}, actively analyzing the detected signals of cosmic rays and individual particles using space telescopes and interpreting observational data in the framework of various scenarios.

The so-called indirect methods of searching for traces of dark matter among numerous astrophysical phenomena simultaneously provide a testing ground for highlighting the most viable options for extending the Standard Model. Having no clues from Nature, we have to sort through the options for the DM construction from the main bricks of matter known to us; fermions, scalars, bosons, compound new hadrons or atoms, neutralinos from supersymmetry, axions, representatives of the Dark World interacting with objects of our world by exchanging special mediators - dark bosons. 

In all cases, we hope to find among the many studied reactions occurring in acts of interaction involving dark matter objects and parts of ordinary matter, unique events that carry information about the dynamics and origin of dark matter. As very important addition we consider events with high-energy neutrino and/or photons that are detected mostly by IceCube and LHAASO (and, certainly, by other ground observatories - because of unpredicted ways of cosmic strangers)\cite{Augier,Hess2,Neutrino1,Neutrino2,NeutrinoAtm,Acc2,Acc3,AccBoost,CR,EUSO,LHAASO}. Note, some interesting data can result from analysis of correlations between observed events with photons and neutrinos from close directions.
 
A special role has recently been played by the analysis of scattering by dark matter objects of particles generated by processes that take place at enormous densities and energies - in particular, in jets from powerful quasars (or blasars), in the vicinity of black holes. Depending on the DM scenario, possible, in principle, various observed signals: monochromatic photons from DM annihilation, high-energy particle fluxes during the decays of a hypothetical supermassive DM, continuous radiation in a certain energy range due to transitions between DM components, and so on.

In this paper, we consider the inelastic interaction of high-energy photons with DM particles. For definiteness, we work within the framework of the SM extension with additional heavy hyperquarks, this is the so-called hypercolor model in its minimal version with two doublets of new fermions and $SU(4)$ symmetry. In this case, it is possible to construct a vector interaction of hyperquark currents with the gauge bosons, providing the necessary smallness of the Pekin-Tackeuchi parameters. Further, in the framework of the llinear sigma model, by analogy with low-energy hadron physics, bound states of hyperfermions are introduced, i.e. new unstable hyperhadrons. On this path, a set of pseudo-Nambu-Goldstone states arises, of which several - the lightest neutral state of a hyperpion triplet and a hyper-diquark with a non-zero conserved hyperdron number - turn out to be stable. These states are interpreted as TM candidates. Some necessary technical details for the description of the process interested will be presented in Section 2. Section 3 contains results of calculations; discussion of results is placed in section Conclusions.

\section{Dark Matter candidates in hypercolor scenario}
Minimal model of the vector hypercolor SM extension contains one doublet of additional heavy H-fermions (H-quarks in confinement) with zero hypercharge. To provide the necessary smallness of Peskin-Takeuchi parameters, initial fields of H-fermions are redefined resulting in Dirac fields that interact vectorially with the gauge fields; extra $SU (2)_w$ symmetry ensures this electroweak interaction. An extra singlet scalar, $\tilde \sigma-$ meson, emerges to provide spontaneous symmetry breaking and, consequently, non-zero masses of new fields. 

At the next stage, using the linear hyper-$\sigma-$ model (as it is done in the low-energy hadron physics), a new H-hadrons generated by H-quarks currents arize with some hierarchy in masses. Besides, the global $SO(4)$ breaking generates a set of pseudo-Nambu-Goldstone (pNG) states, including a triplet of pseudoscalar H-pions and neutral H-baryon (H-diquark with the additive conserving quantum number) along with its antiparticle; H-pions possess a multiplicative conserved quantum number\cite{Bai}. Thus, the neutral states, $\tilde \pi^0$ and $B^0,\, \bar B^0$, are stable in this scenario, so they can be interpreted as the DM candidates with equal masses at the tree level.

Note, the model which is used here to consider high energy photons scattering off the DM is described in detail in a series of papers\cite{OurPRD,OurDM1,OurDM2,WeAdv,OurRev,IJMPA2019}. That is why we do not repeate here all known elements of the scenario; remind only, the DM candidates masses were estimated from analysis of the DM burnout kinetics, and we get: $m_{DM} \sim 1 \, \mbox{TeV}$. The electroweak mass splitting in the H-pions triplet, i.e. between charged and neutral hyperpions, is nearly constant: $\Delta m_{\pi} \approx 0.16\, \mbox{TeV}$. Mass splitting between H-pions and neutral H-baryon, another component of the DM, can be as large as $\approx (10-15) \, \mbox{GeV}$ (see\cite{Bled2021}). Mass of $\tilde \sigma-$ meson is connected with the H-pion mass and depends on the the mixing  angle , $\theta$, between  $\tilde \sigma-$ and Higgs boson. This mixing should be small, $\sin \theta \lesssim 0.1$, so the standard Higgs boson has a small admixture of additional scalar state, $\tilde \sigma$. Note also that the density of H-baryons dominates over density of $\tilde \pi^0$ almost for all possible values of model parameters becaue of different origins of the DM components burning out at different rates\cite{IJMPA2019}. The charged components of H-pion triplet decay, and the dominant channel is the following: $\tilde \pi^{\pm} \to \tilde \pi^0 l^{\pm} \nu_l$ with $\Gamma \approx 3\cdot 10^{-15}\, \mbox {GeV}$. 

 
Certainly, the DM objects are neutral, so they do not interact with photons directly. However, in this scenario there is a class of tree-level diagrams which describe intermediate stage in the total process, $\gamma \tilde \pi^0 \to W^+ \tilde \pi^- \to ll'\nu_l \nu_{l'} \tilde \pi^0$. Here, we know that charged H-pion decays into lepton and neutrino with $Br \approx 1$, and (in fact, intermediate) gauge $W-$ bosons also eventually decay into lepton-neutrino pairs. Correspondingly, we need only in hyperpion triplet properties. From this point of view, this DM component demonstrates possibilities of any WIMP scenario where the DM candidate interact with the standard gauge bosons in some way. Specific detail is: this scenario deal with heavy DM candidates, so to accelerate them the high energy transfer from the projectile is necessary.   

In other words, H-pion as the DM component is the almost “pure” WIMP having tree level elctroweak links to the SM fields due to charged (unstable) components of H-pion triplet.
Indeed, the model contains two types of stable neutral DM candidates, however, scatterings of photons off neutral H-pion are the most simple due to vertices which are shown in the following part of the model Lagrangian:

\begin{equation}
L_{EW}= igW^{\mu}_+(\tilde \pi^0 \tilde \pi^-_{, \mu}-\tilde \pi^0_{, \mu} \tilde \pi^-)-ieA^{\mu}(\tilde \pi^- \tilde \pi^+_{, \mu}-\tilde \pi^-_{, \mu} \tilde \pi^+)+ eg\tilde \pi^0 \tilde \pi^- A^{\mu}W^+_{\mu} + h.c. 
\end{equation} 

The $B^0$ DM component participates in EW interactions only via fermion, vector and/or scalar (gauge boson, quark, H-quark or H-pion) loops and via (pseudo) scalar (Higgs boson,$\tilde \sigma$) exchanges, so, such stable H-diquark presents a hadronic type DM. Some comments on photons scattering off $B^0$, component will be done later.

\section{Neutrino production by photons}


Here, we consider inelastic interaction of photons with energies $\sim (1-10)\, \mbox{TeV}$ with the DM objects. Such process of leptons and neutrinos production by photons seems interesting because the yield of these secondary particles depends both on DM dynamics and distribution its density in space. 

On origin of high-energy photons and neutrinos may be associated, in particular,  with physics and structure of jets from active galaxies nucleus(AGN), events with these particles of $TeV-$ energies are repeatedly observed and analyzed\cite{Plavin,Mauro,Zahir,HESS,Stecker,quasar,Blasars,Shukla,Scatt0,Scatt1,Scatt2,Symmetry2020}. So, there is a possibility for DM structures in the AGN vicinity to participate in the generation of leptons and neutrinos fluxes. Certainly, an analogous process is possible when high-energy photons are scattered by the DM clumps.

Then, in the framework of minimal hypercolor model we study the photons scattering off $\tilde \pi^0-$ component. Such reaction is the most simple, however, calculations of the total cross section with unstable final states were carried out in the continued mass approach(see\cite{Kuksa} and references therein).

Three tree-level diagrams describing the photon scattering off $\tilde \pi^0$ are depicted in Fig.1 Note, there is an additional diagram (see Fig.2) which, however, gives a small contribution to the cross section, because we are interested in dominant configuration with small squared invariant mass of intermediate W-boson. 


Then, the sum of corresponding matrix elements is:
\begin{multline}
M_C+M_{\tilde \pi} +M_W =ige^{\alpha}_{\gamma}e^{\mu}_W(-g_{\mu \alpha}+ \frac{(2k_1-p_2)_{\alpha}(2p_1-k_2)_{\mu}}{d_{\tilde \pi}}+\frac{(p_1+k_1)_{\beta}}{d_W}\cdot\\
(g^{\nu \beta}-\frac{q^{\nu}_W q^{\beta}_W}{m^2_W}(g_{\mu \nu}(p_2-2k_2)_{\alpha}-g_{\nu \alpha}(2p_2-k_2)_{\mu}+g_{\mu \alpha}(p_2+k_2)_{\nu})).
\end{multline}
Here, propagators are denoted as $d_{\tilde \pi}, \, d_W$. 

The squared matrix element is cumbersome, so the exact expression for the cross section of this sub-process is not shown here. Instead, we show  dependencies of differential cross section on squared invariant mass of virtual W-boson, various masses of the DM target and energies of incident photon for fixed energies 
of final H-pion (see Fig.3). Analogous dependencies of differential cross section on squared invariant mass but fixed energies of virtual W-boson and for fixed energy of incident photon are shown in Fig.4. In fact, these two-dimensional curves are sections of three-dimensional graph for different values of the parameters; so, these curves contain information about the angle between products of W-boson decay (lepton and neutrino, in particular).



Of course, it is important to know not only differential cross section for the sub-process with virtual states (unstable charged H-pion and W-boson with known branching of decays into leptons and neutrinos) but total cross section of the process. This next step can be done using model of unstable particles with a smeared masses\cite{Kuksa}. In this approximation the total cross section is presented in a factorized form:
\begin{gather}
\sigma_{tot}(s) = \int d\mu^2 \sigma_{tot}(s,\mu^2) \rho (\mu^2),
\end{gather}
here $\mu^2$ is the variable squared invariant mass of intermediate unstable state and
$$
\rho (\mu^2) =\frac{1}{\pi}\frac{\sqrt{\mu^2}\Gamma(\mu^2)}{((\mu^2-M^2(\mu^2))^2+(\sqrt{\mu^2}\Gamma(\mu^2))^2)^2}
$$
is the density of probability; integration goes over kinematically allowed region.
Now, the total cross section for the whole process can be estimated with a good accuracy (note, the approach above was tested in a numerous reactions, and the accuracy was reliably evaluated as $\lesssim 5\,\%$).

Total cross section for the process of photons scattering off scalar DM candidates, $\gamma \tilde \pi^0 \to W^+ \tilde \pi^- \to ll'\nu_l \nu_{l'} \tilde \pi^0$, in dependence of photons energy are presented in Fig.5. Here we consider not very high energies of photons, up to $10 \, \mbox{TeV}$ because for higher energies the photon flux is much lower. These results depend on the DM mass weakly, so we use here some referent value, $m_{DM} =1\, \mbox{TeV}$. Besides, we get the cross section almost stable in the energy region considered. This result indicates some possible correlations between detected high-energy photons and neutrinos fluxes - namely, neutrinos of high energy can be produced by photons scattered on the DM objects.

But this is not the whole story because there are loop contributions to neutrino+ leptons generation by the photons scattering off another DM component, stable H-baryon. 
However, this DM candidate presents other type of DM objects which do not interact with the gauge boson directly. So, one from possible contributions into such type reaction is shown in Fig.6. However, there should be also additional channels to transform a part of high-energy photons into fluxes of leptons and neutrinos. These are processes like $\gamma B^0 \to W^+ W^- B^0, \,\, \gamma B^0 \to t \bar t B^0,\,\, \gamma B^0 \to Z B^0$ with subsequent decays of heavy standard quarks and gauge bosons. 
These interesting and promising channels of inelastic interactions of photons with the DM will be analyzed elsewhere.

\begin{figure}
	\centering{\includegraphics[width=10 cm]{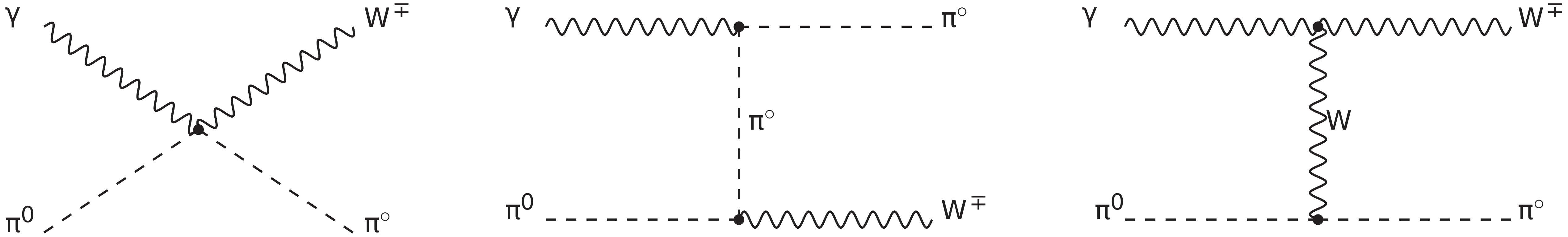}}
	\caption{Tree-level diagrams for the subprocess $\gamma \tilde \pi^0 \to W^{\pm} \tilde \pi^{\pm}$.}
	\label{fig_1}
\end{figure}

\begin{figure}
	\centering{\includegraphics[width=3 cm]{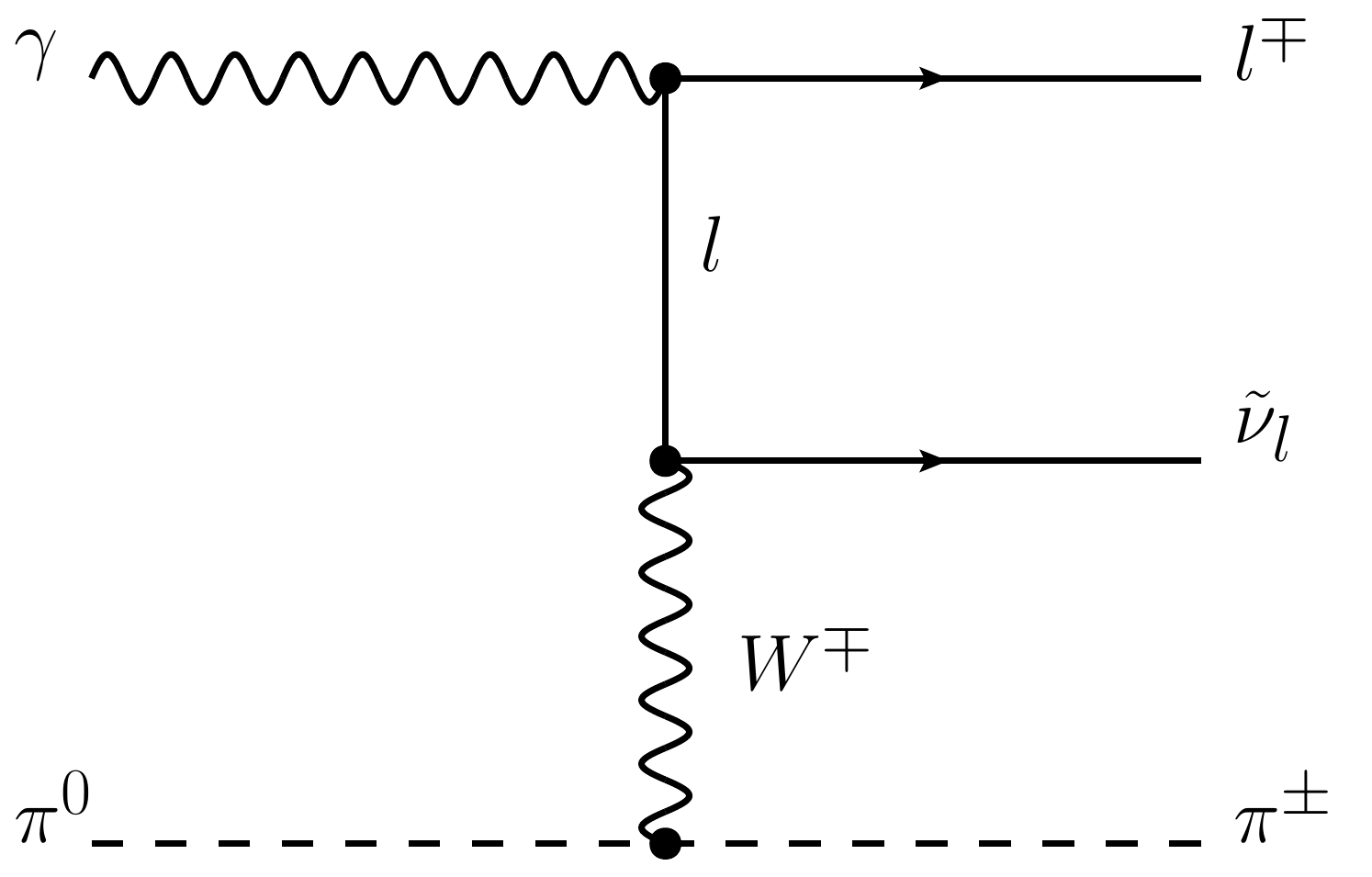}}
	\caption{Additional contribution to the scattering process.}
	\label{fig_2}
\end{figure}

\begin{figure}[h!]
	\centering
	\begin{minipage}[b]{0.45\textwidth}
		{\includegraphics[width=\textwidth]{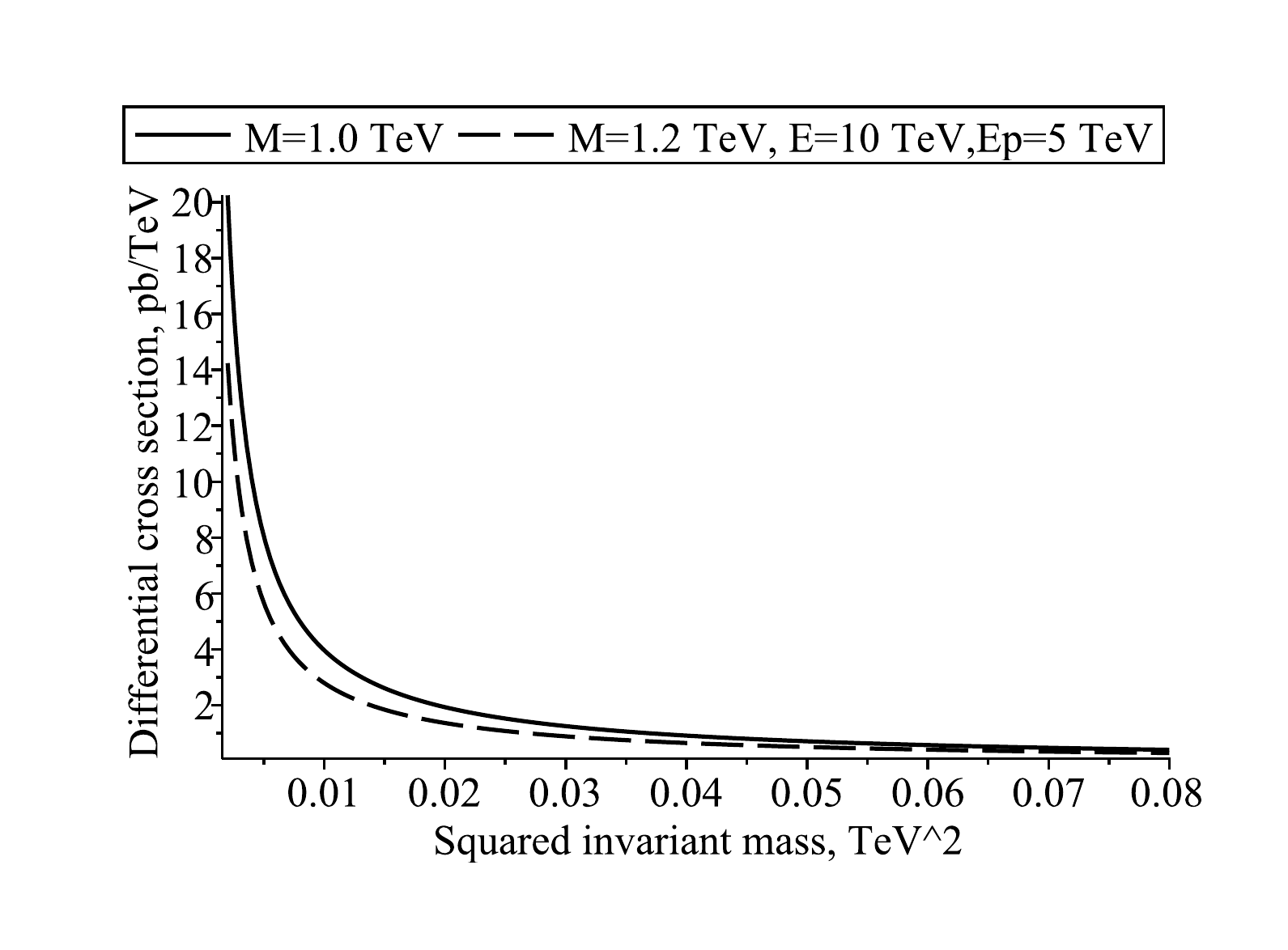} a)}
	\end{minipage}
	\begin{minipage}[b]{0.45\textwidth}
		{\includegraphics[width=\textwidth]{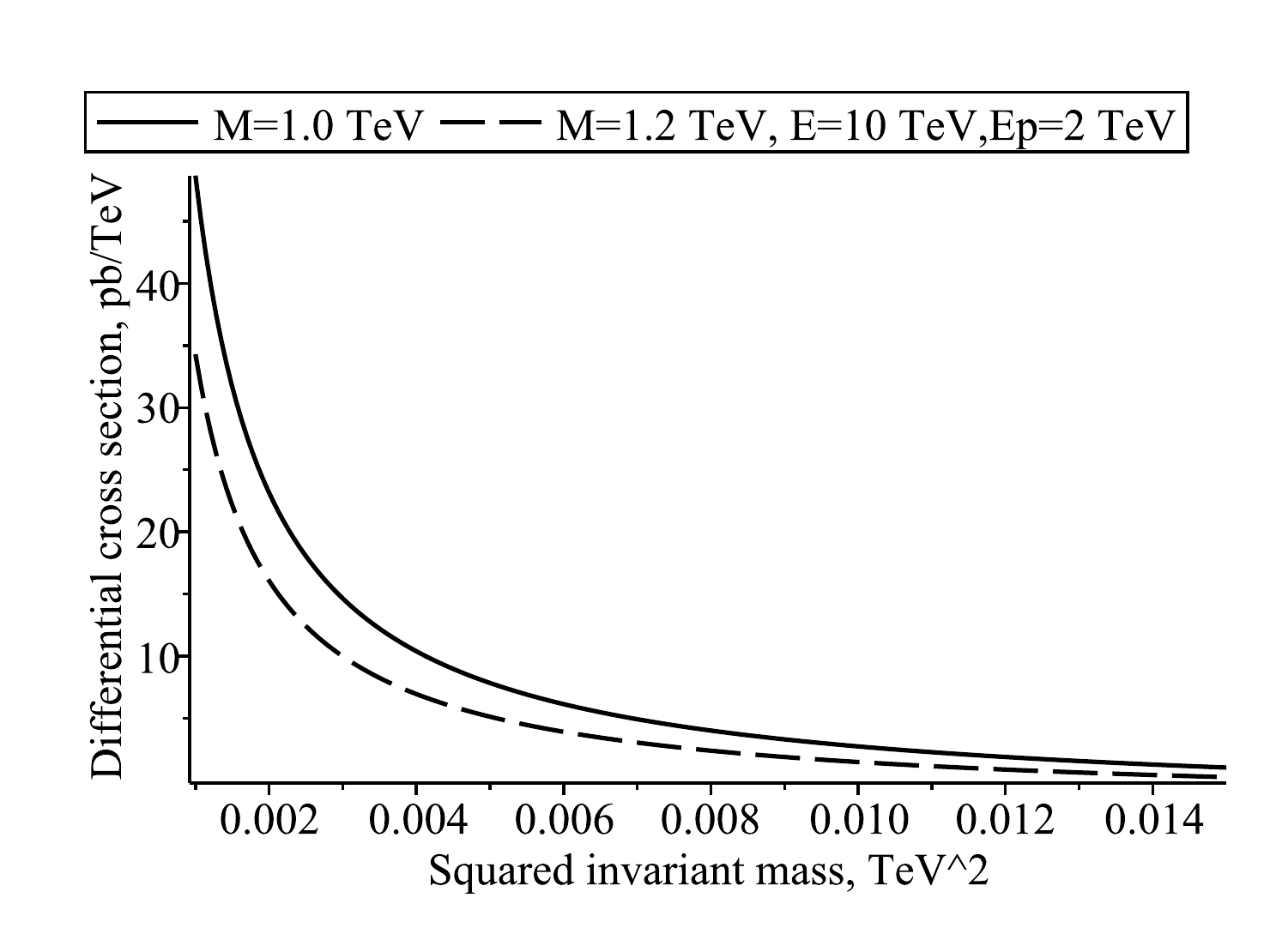} b)}
	\end{minipage}
	\caption{Differential cross section in dependence on W-boson squared invariant mass for the incident photon energy $E_{\gamma} = 10 \, \mbox{TeV}$ with the fixed energy of final H-pion: a) $E_{\tilde \pi}=2\, \mbox{TeV}$; b) $E_{\tilde \pi}=5\, \mbox{TeV}$.}
	\label{fig_3}
\end{figure}

\begin{figure}[h!]
	\centering
	\begin{minipage}[b]{0.45\textwidth}
		{\includegraphics[width=\textwidth]{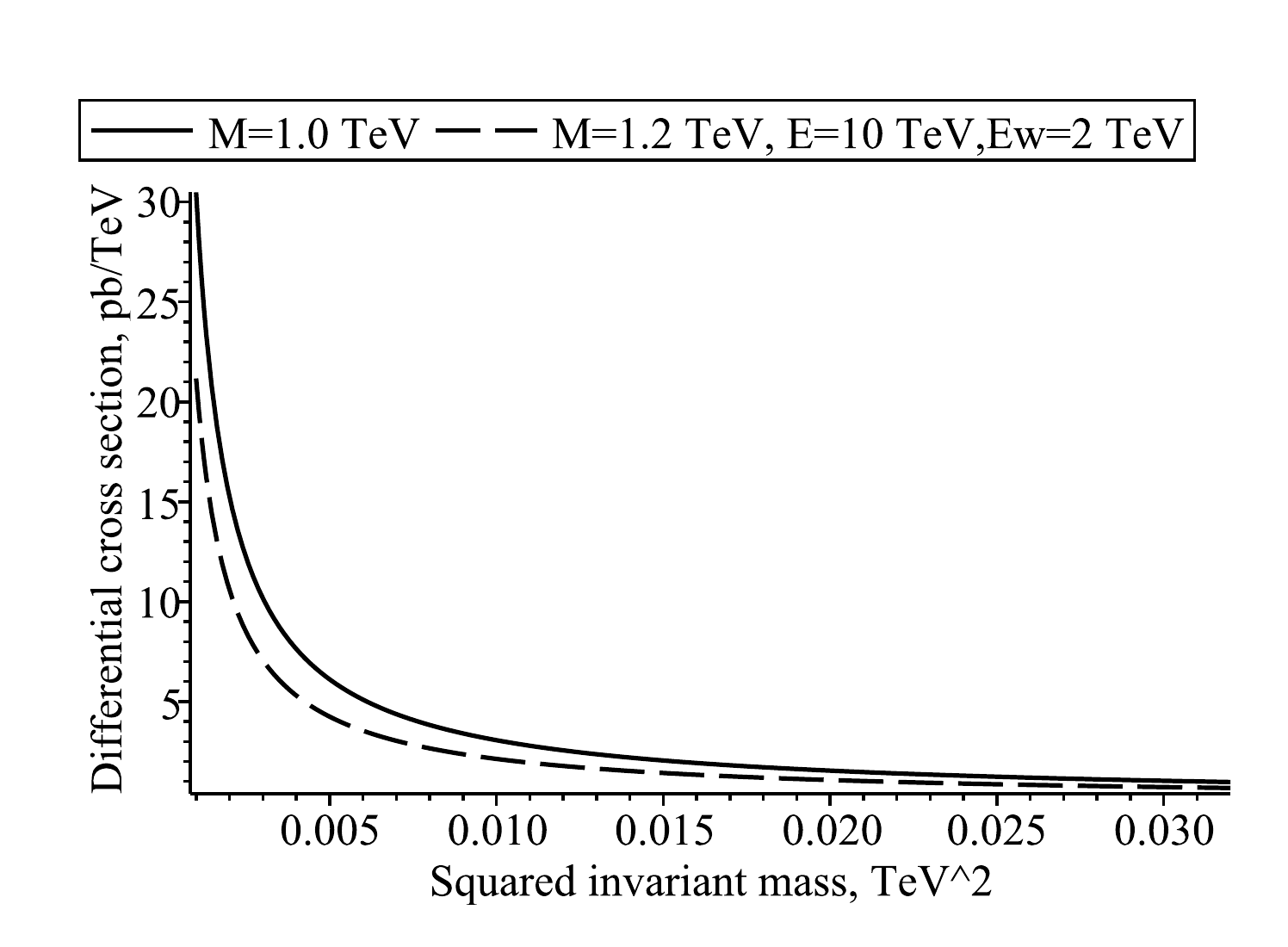} a)}
	\end{minipage}
	\begin{minipage}[b]{0.45\textwidth}
		{\includegraphics[width=\textwidth]{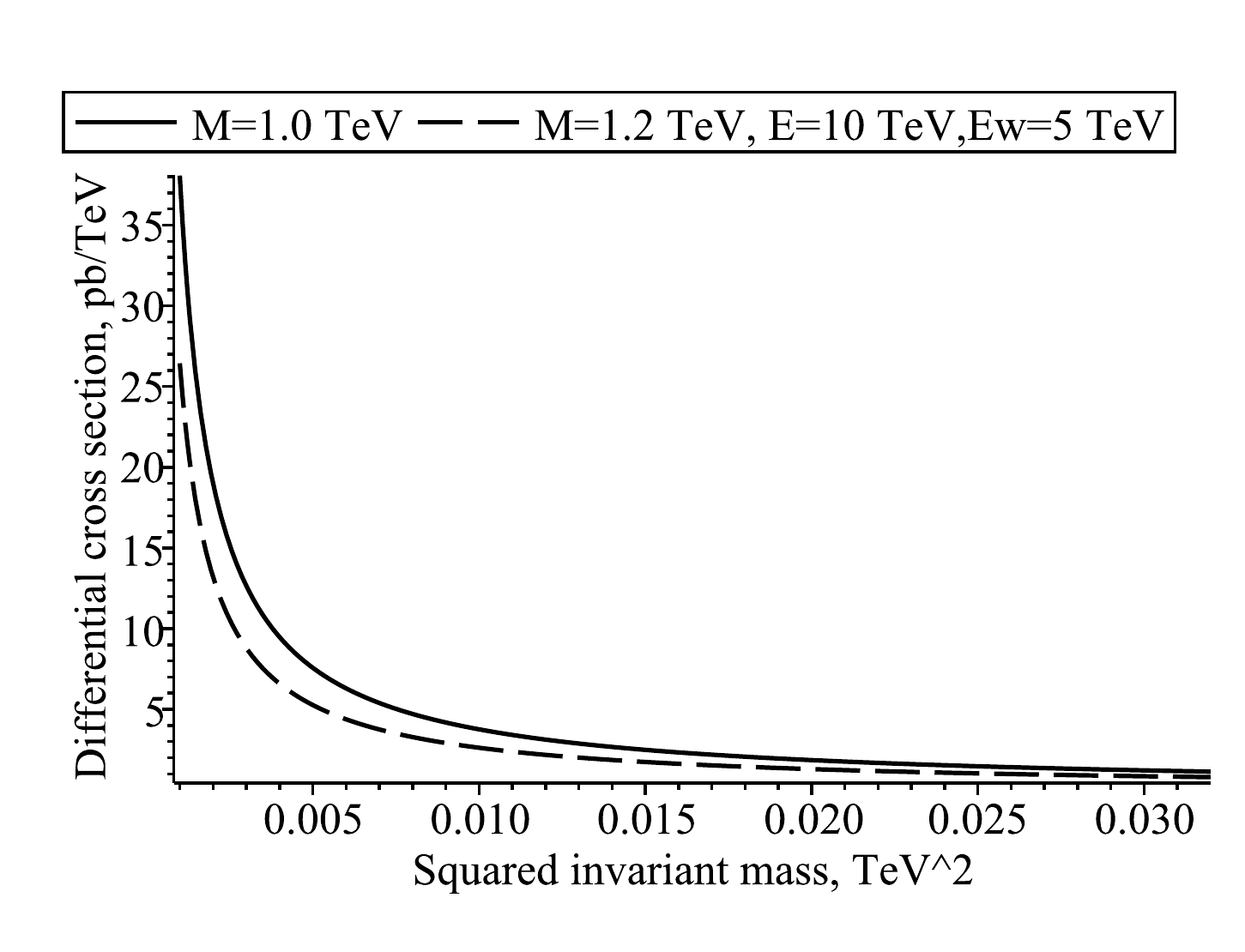} b)}
	\end{minipage}
	\caption{Differential cross section in dependence on W-boson squared invariant mass for the incident photon energy $E_{\gamma} = 10 \, \mbox{TeV}$ with the fixed energy of W-boson: a) $E_{W}=2\, \mbox{TeV}$; b) $E_{W}=5\, \mbox{TeV}$.}
	\label{fig_4}
\end{figure}

\begin{figure}[h!]
	\centering
	\begin{minipage}[b]{0.45\textwidth}
		{\includegraphics[width=\textwidth]{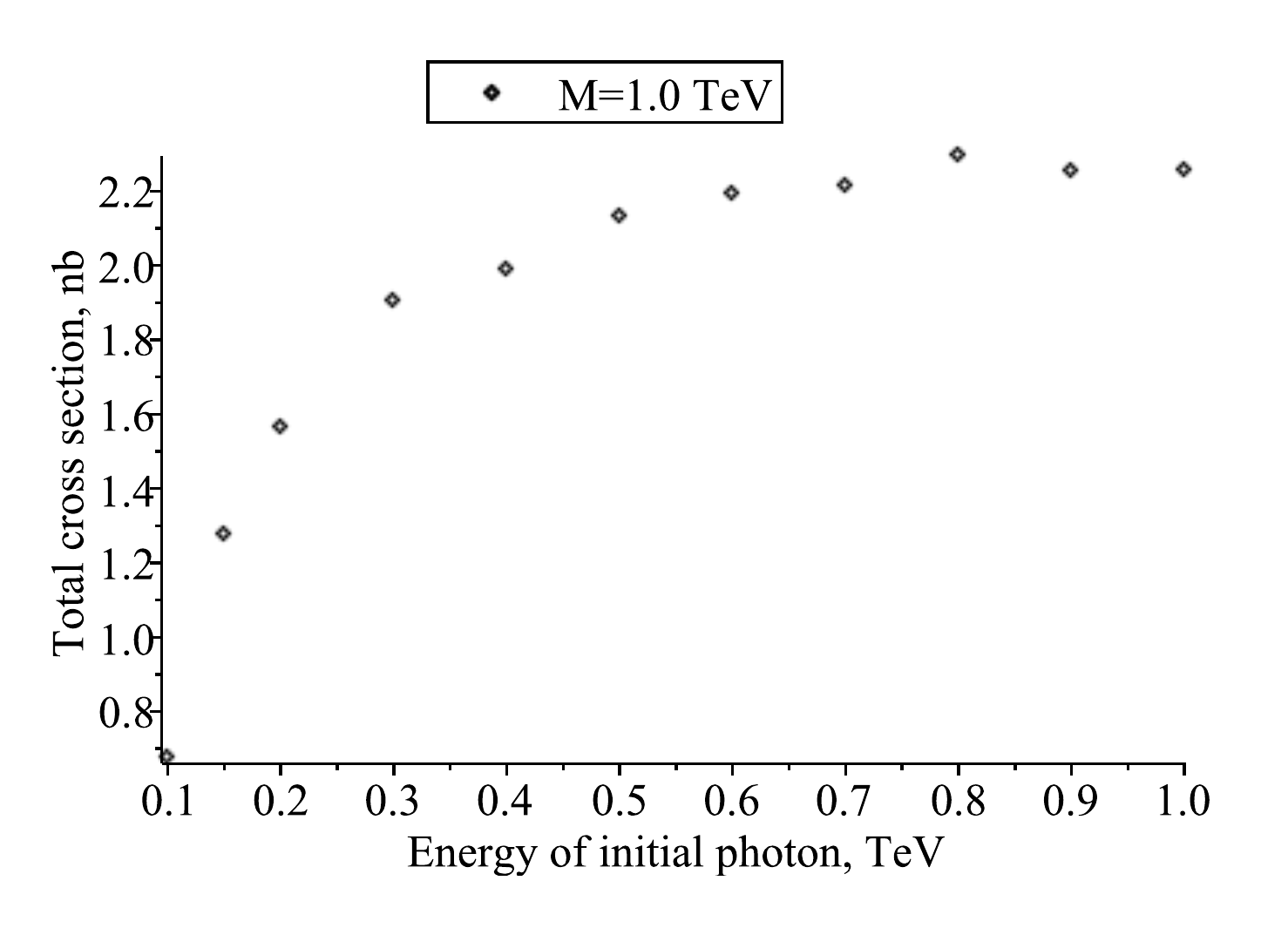} a)}
	\end{minipage}
	\begin{minipage}[b]{0.45\textwidth}
		{\includegraphics[width=\textwidth]{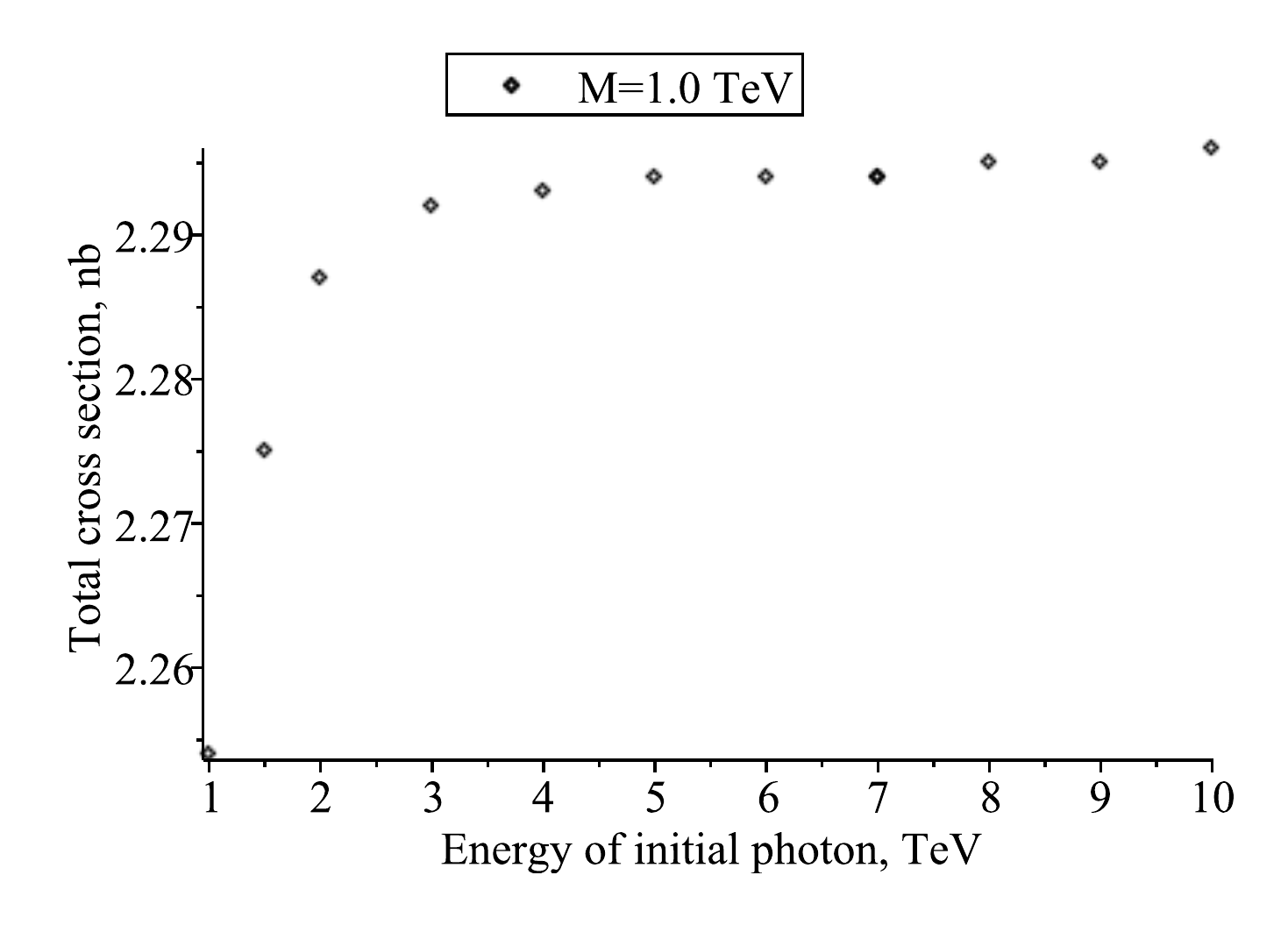} b)}
	\end{minipage}
	\caption{Total cross section in dependence on photon energies energy; of incident photon: a) up to $E_{\gamma} = 1 \, \mbox{TeV}$; b) up to $E_{\gamma} = 10 \, \mbox{TeV}$.}
	\label{fig_5}
\end{figure}

\begin{figure}
	\centering{\includegraphics[width=3 cm]{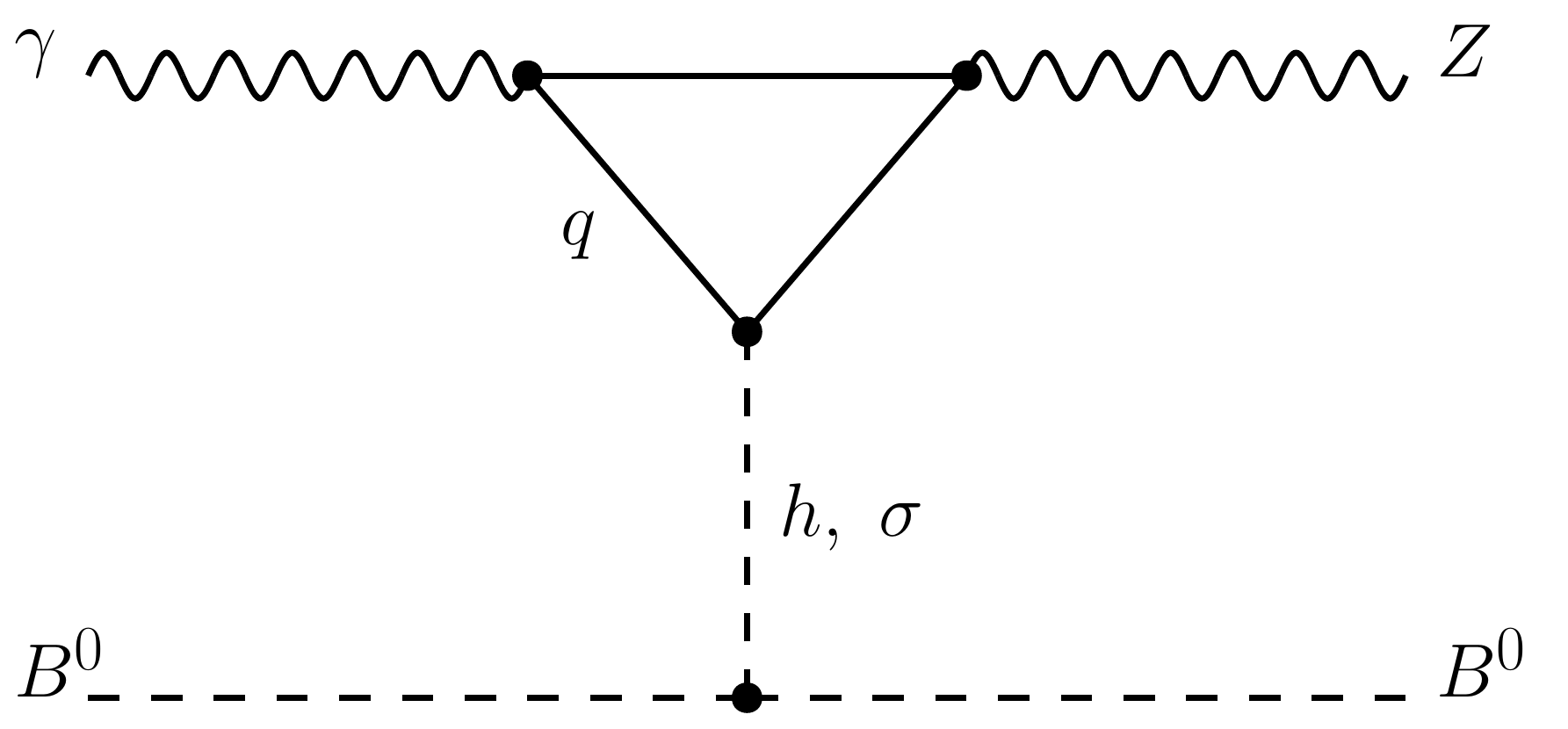}}
	\caption{One from possible loop-level diagrams for the subprocess $\gamma B^0 \to Z B^0$.}
	\label{fig_6}
\end{figure}

\section{Conclusions and some open questions} 
 
So, for tree level production of leptons and neutrinos by inelastic scattering of high-energy photons off the DM we get $\sigma_{tot}(E_{\gamma}) \sim 10^0 \, \mbox{nb}$. 

As it is seen from calculations, a significant part of $E_{\gamma}$ is converted into energies of secondary leptons and
neutrinos which are generated by decays of $W-$boson both in direct channel and from unstable light standard mesons due to hadronic decay channels of W.

At the time, a DM target is an active and necessary participant of the process, so, it also can get sufficiently large portion of photon energy, so the DM components can be accelerated up to energies $\sim (1-10)\, \mbox{TeV}$ or even larger depending on $E_{\gamma}$. Intensity of this reaction strongly depends on the DM density (the macroscopic cross section is $\sim \rho_{DM}$, the high-enery secondaries yield can be sufficiently enhanced when the scattering occurs in regions of high DM density.

It is an important notification because the main sources of high-energy photons are quasars(or blasars shining in the direction of the Earth) which emit dense and very fast fluxes of charged particles as jets (see references above); high-energy photons are radiated by these charged particles, so photons and neutrinos generated in quasars and blasars jets can be detected by ground observatories and cosmic telescopes. Inelastic scattering of photons off the DM most effectively occurs in the vicinity of quasars, i.e. close to active nuclei of galaxies. And it is in these regions of strongest gravity that the DM density is the largest.

Products of this collision of photon with the DM are approximately collimated with the initial photon direction for high $E_{\gamma}$. 
Flux of neutrinos produced with cross section $\sim (0.1-1)\, \mbox{nb}$ and energies $\sim (1-10)\, \mbox{TeV}$ should correlate with the initial photon flux, consequently, signals of TeV-photons detected by cosmic telescopes and ground observatories can be (approximately) synchronized with neutrinos of close energy which come from the nearly the same direction.

Possibly, interactions of such type affect on the DM density fluctuations at early stage of evolution when cosmological plasma is hot and contains a lot of the DM (neutral) objects and also the dense fluxes of photon radiaton. So, these processes can also affect on the density of photons and DM carriers throughot the radiation dominated era. In some sense, these processes can somewhat wash out dense DM clumps and fluctuations due to accelerating the DM particles of various masses.

It is, of course, a hard task to find out and separate from any other sources neutrinos and photons signals correlated in time and spatial direction, however, if it were discovered, it would mean obtaining an important information about the structure and dynamics of the DM and blasars jets, as well as about the DM spatial distribution and its ability to affect to dynamics and composition of cosmic rays in space.



In any case, it is worth considering such processes of photons transformation into leptons and neutrinos fluxes accompanying with the DM accelerated particles in all possible DM scenarios and for various characteristics of cosmological evolution stages. The DM plays an important role of an active and necessary catalyst of mutual transitions between various types of the SM particles.
As it seems, such processes with the obligatory presence of the DM should be taken into acount when the early Universe dynamics and structure are studied.

\section*{Acknowledgements}
The work has been supported by the grant of the Russian Science Foundation No-18-12-00213-P https://rscf.ru/project/18-12-00213/
and performed in Southern Federal University.


\end{document}